# Nanoscale symmetry protection of the reciprocal acoustoelectric effect

Sandeep Vijayan[1], Stephan Suffit[1], Scott E. Cooper[1], Yejun Feng[1,*]

[1]Okinawa Institute of Science and Technology Graduate University, Onna, Okinawa 904-0495, Japan
*Corresponding author: yejun@oist.jp

## ABSTRACT:

**Neumann's principle states that all physical properties of a material are bound by its symmetry. While bulk crystals follow well-defined point and space groups, phenomena at a substrate's surface could have less apparent symmetry origins. Here we experimentally explore both reciprocal and non-reciprocal types of acoustoelectric (AE) effects driven by surface acoustic waves (SAW). The non-reciprocal AE voltage is connected to the natural single-phase unidirectional transducer from device engineering. On the other hand, reciprocal AE effect exists in certain SAW configurations that are of different symmetry origins. Half of the configurations have a valid reciprocity-preserving symmetry element of either a mirror plane or an even-order rotational axis that is perpendicular to the substrate surface. The other half of the configurations do not possess reciprocity-preserving symmetry operations of the half-space but have the SAW propagation and the surface normal directions interchanged from the first scenario. Here, the reciprocity of SAW states is protected by the symmetric structure of the nanoscale strain tensor. The correspondence between two types of configurations has its origin imbedded in the SAW composition of both compression and shear waves along two orthogonal directions respectively.**

**Keywords:**
Acoustoelectric effect; surface acoustic waves; non-reciprocity; NSPUDT; mirror and rotational symmetries.

## Introduction

Modern science and technology rely heavily on the dispersion relationship of electromagnetic waves. From audio-frequency radiations to optical lights and x-rays, the relationship between the characteristic length scale and energy is governed by the linear dispersion condition of light, $E = \hbar ck = hc/\lambda$, with $\lambda$ the wavelength, $E$ the energy, $k = 2\pi/\lambda$ the wavevector, $h$ the Planck constant, and $c$ the speed of light. On the other hand, acoustic waves also offer a linear dispersion but with a much smaller ratio $dE/dk$ as the dispersion is determined by the speed of sound (in solids) instead of the speed of light, as they are separated by five orders of magnitude. On the surface of a substrate, acoustic waves can form and propagate but extend no deeper than one wavelength into the bulk [1, 2], which makes surface acoustic waves (SAW) a perfect choice to stimulate and probe physical systems in a confined space. With modern nanofabrication technology, this unique dispersion allows experimental coupling to physical phenomena at length scales varying from nanoscale (~10 nm), to mesoscopic, to hydrodynamic (~100μm), while using much subtler perturbation energies than those of electromagnetic waves. For example, SAW has demonstrated an extremely efficient coupling to the non-Abelian type, even-denominator $\nu = 1/2$ fractional quantum Hall state in a two-dimensional electron gas [3].

SAW techniques have been widely utilized to explore many emergent frontiers of physical science, ranging from quantum information manipulation [4] to single electron circuitry [5], battery research [6], the creation and transportation of solitonic skyrmions [7], and microfluids [8]. The microwave driven SAW environment is often complemented by other measurement techniques of electrical [5, 6] and optical [7] nature. In recent decades, there has been a fast-growing effort to utilize acoustoelectric (AE) effect [9, 10] to probe spintronic topics [11-14]. Among all studies, there exist many accounts of asymmetric or non-reciprocal acoustoelectric behavior, which demonstrate different amplitudes when they are driven by oppositely traveling SAW states [11-16]. Most of these studies involve magnetic materials, so the non-reciprocal property is often attributed to ferromagnetic sample films that possess a broken time reversal symmetry [11-14], or the metal-insulator transition that is concurrent with the paramagnetic-ferromagnetic transition [15]. In some cases, when two interdigital transducers are used, the asymmetry can be attributed to inequivalent transduction efficiencies of the interdigital transducers (IDT) [16]. As research efforts are pushing towards measurements' sensitivity limit, a fine control of the device performance and a complete understanding of the non-reciprocal AE effect becomes necessary and imperative.

In the literature of SAW device engineering, asymmetric phenomena have been reported in one special type of devices named natural single-phase unidirectional transducer (NSPUDT) even though electrodes are symmetrically placed [2, 17-20]. This phenomenon was attributed to the influence from the mass of IDT electrodes on the reflectivity coefficient $r^{(\pm)} = R_e^{(\pm)} + R_m^{(\pm)}\, h_{\text{metal}}/\lambda$ of SAW states, with $R_e^{(\pm)}$ and $R_m^{(\pm)}$ representing the electrical and mechanical effects respectively, $(\pm)$ of two opposite propagation directions, $h_{\text{metal}}$ the height of metal electrodes, and $\lambda$ the SAW wavelength [2]. $R_e^{(\pm)}$ are regarded as always symmetric, while $R_m^{(\pm)}$ can be asymmetric. Based on the perspective of reflection center position, Ref. [18] evaluated a limited number of high symmetry directions among 12 monaxial point groups (out of a total of 13 except type *1*) but did not clarify the key symmetry reason of NSPUDT's existence. Using numerical coupling-of-modes simulations that also includes the reflection effects, Ref. [19] explored the conditions of pure standing wave state which also address the NSPUDT effect. Nevertheless, there remains a lack of clear symmetry logic amid misidentified symmetry elements (see below). Overall, the general NSPUDT effect was considered uncommon in the later literature [2] and was specifically believed not affecting SAW propagation along high symmetry directions [2]. The unclear situation caused a subsequent study [20] to make unnecessary discussions about the three-fold symmetry in $LiNbO_3$ and assign the substrate surface plane with the mirror symmetry.

Here working with non-magnetic conductors, we have searched for and discovered non-reciprocal AE effects that can be experimentally attributed to NSPUDT-type of SAW states. Working with the most popular SAW substrates $LiNbO_3$ and $LiTaO_3$ of the *3m* point group, we experimentally resolve the configurations that separate the reciprocal and non-reciprocal AE effects. While theoretical SAW states constructed from eigenstate solutions of Stroh's tensor equation of motion would always satisfy the time reversal symmetry, in real devices, multiple reflections of propagating SAW states inside the IDT break the time reversal symmetry and can introduce the non-reciprocal AE effect. Configurations leading to the reciprocal AE effect are limited to two scenarios. In $LiNbO_3$/$LiTaO_3$ based devices, the first scenario has a SAW state propagating perpendicular to a mirror plane of the half-space, which can be extended to all point groups with an alternative symmetry operation of an $180^o$-rotation around the substrate surface normal; here the essence of scenario one is that either of these two global symmetry operations can directly connect two oppositely propagating SAW states. The second scenario has a one-to-one correspondence to the first scenario, but with directions of the surface normal $\hat{\boldsymbol{n}}$ and propagation $\hat{\boldsymbol{m}}$ interchanged. Despite a lack of global symmetry, the second scenario has its relationship to the first scenario sustained by the symmetric strain tensor over the nanometer scale. This microscopic origin can be regarded as a hidden symmetry that protects reciprocal SAW states. All other cases lead to non-reciprocal SAW states and AE effect. Our experimental results and identified cause of strain tensor symmetry are critical to understand both reciprocal and non-reciprocal types of AE effects and are related to pure standing waves in real devices. Our general symmetry guideline should influence many experimental studies of acoustoelectric, spintronic, and quantum transport nature.

**Results**

**High-sensitivity measurement of the acoustoelectric effect**

Many past studies of the AE effect have used DC techniques that can be highly susceptible to the environment condition [11, 12]. Here we have developed an AC lock-in measurement technique to detect the AE effect in high fidelity (Methods). The efficiency and high sensitivity of our measurement scheme allows a fine survey of the full spectral shape, leading to a precise understanding of mechanisms contributing to the SAW state as we demonstrate below. In the simplest scenario, a surface acoustic wave is expected to have a Sinc-squared functional form [1, 2]. In our measurement (Fig. 1b), signals of this form can be clearly observed over four decades of dynamic range, exemplified by a device with a low number of IDT pairs $N$ (~20). From the Sinc-squared functional form, we can extract $N = 19.3 \pm 2.4$, consistent with the design value of 20. The small $N$ value ensures that the generated SAW experiences limited internal reflections at IDT fingers before it escapes outside the IDT, as the negligence of reflection is implicitly assumed in the derivation of Sinc-square form [1, 2].

When the pair number $N$ of the IDT is increased to a large number, the spectral shape evolves away from the Sinc-squared form. Here, even a small internal reflection at each metallic finger of the IDT can have an accumulated effect within the large array and change the spectral form to Lorentzian [21]. As a SAW's frequency varies, its wavelength slightly differs from the IDT periodicity. When the total number of pairs is small, a slightly different wavelength still allows the SAW state to be coherent across the full range of IDT, leading to the Sinc-squared spectral form. When $N$ is large, the increasing internal reflection causes the coherence length of SAW becomes shorter than the length of IDT, leading to a Lorentzian spectral form [21]. A Lorentzian fitting to the data in Fig. 1c gives the FWHM of $3.04 \pm 0.08$ MHz, that is clearly larger than what is expected for the 3dB width of a $N = 250$ IDT device with the Sinc-squared form (~1.78 MHz) [1]. The whole length of IDT remains statistically sensible to the SAW states, as there exist fast oscillations superposed on the overall Lorentzian form with a period about $1.85 \pm 0.10$ MHz, suggesting an estimated $N = 267 \pm 15$ that is consistent with our design parameters.

**Non-reciprocal AE effect and SAW states**

With the demonstration that the AE effect can precisely reflect the propagating SAW, we now probe the symmetry of SAW states along opposite directions. Two oppositely traveling SAW states can be regarded as reciprocal to each other if they can be related by the time reversal symmetry operation. Studies in the literature often used a configuration of two in-line but distanced IDTs with one patch of metallic film in between, so each IDT can drive a SAW state to the same AE probe [11, 15]. Here we detect the AE effect on two sides of a single IDT [17] (Fig. 1a), using two identical patches of metallic films at equal distance away from the IDT (Methods). This configuration ensures that the oppositely propagating SAWs are created under the same transduction process.

In the configuration of $x$-axis propagation on a $128YX$-cut substrate, AE measurements along forward $(+x)$ and backward $(-x)$ directions generate identical frequency profiles (Fig. 2), which verifies the general scheme of our measurements and our devices' reliability. This reciprocity of SAW along $\pm x$ directions is robust that when we change the IDT parameters such as $N$ and/or the metallic finger height $h$, only the frequency spectrum's magnitude and shape (from Sinc-squared to Lorentzian) vary, but the reciprocal relationship is maintained between oppositely propagating SAWs (Fig. 2).

On the other hand, when SAW is propagating along certain directions, such as the in-plane $Y'$ direction of the $128YX$-cut substrate (Fig. 3), our measured AE voltages $V_{\mathrm{AE}}$ from oppositely propagating SAWs demonstrate a significant difference in their amplitudes, well beyond measurement uncertainty. The $V_{\mathrm{AE}}$ difference in Fig. 3 is dependent on parameters of the IDT structure. For example, an increased thickness $h_{\mathrm{metal}}$ of metallic fingers would enhance this asymmetry (Fig. 3). For devices with a low number of fingers ($N \sim 20$), the spectral difference resides at the central frequency. However, the difference is shifted sideways when $N$ becomes large, and the line shape emerges as two Lorentzian peaks separated in frequency; the $V_{\mathrm{AE}}$ difference is more prominent in one (lower) peak of the two (Fig. 3).

In the literature, measured non-reciprocal AE effect has often been attributed to properties of the thin film under study [11, 20]. Here, we have used trivial, non-magnetic and polycrystalline metal films that are identical for different reciprocal behaviors in both Figs. 2 and 3. Therefore, there is a clear scenario of non-reciprocal surface acoustic waves because of geometrical configurations relative to the substrate, which in turn leads to non-reciprocal AE effect.

A traveling SAW state necessarily satisfies the time reversal symmetry as a charge phenomenon (Supplementary Materials). The non-reciprocity between strengths of oppositely propagating SAW states thus originates from the stage where SAWs are generated at the interdigital transducer towards both directions, with multiple scattering an integral component of the process. The positive correlation between spectral weight difference and loaded mass (Fig. 3) confirms the mechanical reflection's influence on the non-reciprocity. The experimental non-reciprocal AE effect thus is closely related to the NSPUDT effect from the SAW device literature.

**Symmetry elements of reciprocal SAW**

Here we focus on two SAW substrate materials of $LiNbO_3$ and $LiTaO_3$ crystals to experimentally examine the relationship between substrate choices and reciprocal/non-reciprocal AE effects, as they are among the most utilized substrates in both the literature and applications. In the piezoelectric phase at room temperature, these crystals belong to the same symmetry group ($R$3c, #161), with non-centrosymmetric point group *3m* operations solely generated by a mirror $y$-$z$ plane and a threefold rotation axis along the $z$ direction. Most of the devices listed in Fig. 4 have a large $N$ =250 and a large mass load on electrodes, in order to enhance the existence of non-reciprocity. Through Figs. 2 to 4, a total of ten different configurations are evaluated for both $LiNbO_3$ and $LiTaO_3$. Among them, there are five cases of reciprocal and non-reciprocal AE effects respectively.

We note that all cases of reciprocal AE effects in Figs. 2 and 4 can be attributed to either of the two scenarios: (1) the SAW propagation direction $\hat{\boldsymbol{m}}$ is perpendicular to a mirror plane of the half-space (*e.g.* along the $x$-axis when the axis is within the surface), regardless of the surface normal direction $\hat{\boldsymbol{n}}$ of the substrate, and (2) the substrate surface normal $\hat{\boldsymbol{n}}$ and propagation direction $\hat{\boldsymbol{m}}$ are interchanged from configurations of scenario (1). When $\hat{\boldsymbol{n}}$ is parallel to an axis such as the $x$-axis, the propagation direction $\hat{\boldsymbol{m}}$ can be along any direction within the substrate surface, *e.g.* the $y$-$z$ plane. The scenario (1) is justified as two oppositely propagating SAW states can be related by the mirror plane as images of each other; the mirror operation remains a proper symmetry element for the half space. On the other hand, for the second scenario, there does not exist any symmetry operation that directly relates the two oppositely propagating SAW states.

The second scenario appears random and was uncharacterized in Refs. [2, 18]. Practically, there exist only a few stable SAW propagation directions without beam steering to make real devices with the $x$-cut substrates of both $LiNbO_3$ and $LiTaO_3$ [22, 23]. However, configurations of scenario (2) have a one-to-one correspondence to configurations in scenario (1), and the link provides the missing piece to fully understand the symmetry condition for reciprocal SAW states. We note that a SAW state can be constructed from eigenstate solutions of an 8×8 Stroh's tensor equation $\aleph\xi = p\boldsymbol{\xi}$ [19, 24, 25] (Supplementary Materials). Here the $\aleph$ matrix is fully specified by the vectors $\hat{\boldsymbol{m}}$ and $\hat{\boldsymbol{n}}$, three property tensors of elastic constants $c_{ijkl}$, piezoelectric constants $e_{ijk}$, and dielectric permittivity $\varepsilon_{ij}$ of the substrate crystal. The eight-component eigenvectors $\boldsymbol{\xi}^T = (\boldsymbol{A}, \phi, \boldsymbol{L}, D_n)$ are composed of the displacement vector $\boldsymbol{A}$, the traction vector $\boldsymbol{L}$, the electrical potential $\phi$, and the normal component of the electric displacement field $D_n$. $p_\alpha$ ($\alpha = 1,2,\dots,8$) are eigenvalues. The 8×8 tensor equation can be transformed into an eigenvalue equation of $p_\alpha$ (Equation S7 of the Supplementary Materials), in which $\hat{\boldsymbol{m}}$ and $\hat{\boldsymbol{n}}$ appear as structurally symmetric components. Interchanging $\hat{\boldsymbol{m}}$ and $\hat{\boldsymbol{n}}$ would cause differently valued matrices, but two matrices still have the same matrix shape as of zero and non-zero values (Supplementary Materials). So the one-to-one correspondence in choices of ($\hat{\boldsymbol{m}}$, $\hat{\boldsymbol{n}}$) between scenarios (1) and (2) is rooted in the symmetry of ℵ matrix equation for the ($\hat{\boldsymbol{m}}$, $\hat{\boldsymbol{n}}$) pair and goes well beyond the point group symmetry-based arguments in Refs. [2, 18]. Indeed, the ($\hat{\boldsymbol{m}}$, $\hat{\boldsymbol{n}}$) exchangeability in the ℵ matrix equation acts like a hidden symmetry to protect our scenario (2).

All other cases beyond these two scenarios (Figs. 3 and 4a-d), as long as they do not have either $\hat{\boldsymbol{m}}$ or $\hat{\boldsymbol{n}}$ perpendicular to a mirror plane, lead to non-reciprocal SAWs and subsequently asymmetric acoustoelectric effects. Our experimental results of reciprocal AE effects in $LiNbO_3$ and $LiTaO_3$ match with theoretical cases (1) and (3) described in Eq. 24 of Ref. [19], although the emphasis of symmetry differ. Eq. 24 of Ref. [19] also listed another two cases (2) and (4) for substrates such as quartz of point group *32*. These two cases also have a one-to-one correspondence that reflects the symmetry of ($\hat{\boldsymbol{m}}$, $\hat{\boldsymbol{n}}$) in the eigenvalue equation S7 of our Supplementary Materials. Here, the symmetry element is an even-fold rotation axis, when it is perpendicular to the substrate surface. However, when the axis is parallel to the surface, the even-fold axis is no longer a good symmetry element. We again have a situation that the symmetry of Stroh's equation protects the SAW reciprocity in case (2) of Ref. [19] instead of a lattice symmetry element of the point group.

With our understanding of the reciprocity condition in $LiNbO_3$, it is possible to reexamine reported non-reciprocal cases in the literature. In Ref. [15], it was clear that the SAW states propagated along the $z$-axis of $Y$-cut $LiNbO_3$, which is a direction one would expect non-reciprocal SAW states in the substrate. Their understanding of (La,Ca)$MnO_3$ thus likely requires a re-examination. On the other hand, in Ref. [16], with a substrate of 128°$YX$-cut $LiNbO_3$ and a SAW speed of 3980 m/s, the propagation direction was along the $x$-axis [26]. The observed non-reciprocity should be due to device imperfection and other sample distance effects as the authors stated [16].

**Direct measurement of the non-reciprocal AE effect**

Using a configuration of a thin metal film placed in between two IDTs (Fig. 5 schematic), our AC-lockin technique (Methods) can directly extract the non-reciprocal AE effect. While the scheme listed in Fig.1 probes the AE effect using the primary harmonic, we now inject AC modulated radio frequency signals alternatively on two IDTs to excite oscillating SAW states in the substrate and measure both the first and second harmonic components of the acoustoelectric response in the thin film (Fig. 5). As alternatively propagating SAWs create oscillating acoustoelectric potentials, the AC signal at the primary frequency represents the averaged AE effect. The non-reciprocal part of the AE effect can be regarded as a unidirectional voltage; its modulation now follows the second harmonic frequency (Fig. 5a).

In Fig. 5b, we demonstrate both the primary and second harmonic measurements of the AE voltage. The exemplifying device has both a high IDT pair number $N$=250, and a large electrode mass load $h_{\mathrm{Al}}$ = 100 nm, which leads to a double-Lorentzian profile in the first harmonic AE signal $V_{1\mathrm{F}}$ (red curve in Fig. 5b). We choose the geometry of the fabricated device with SAW propagating along the $y$-axis of a $z$-cut $LiNbO_3$ substrate. This provides a comparison between the current measurement to the case study in Fig 4a of a single IDT and two acoustoelectric measurement pads. $V_{1\mathrm{F}}$ measured in Fig. 5b is very similar to the average of two oppositely traveling curves in Fig. 4a, except a few extra oscillations on top of the spectral weight at 480 MHz; these are triple-pass SAW interference patterns between two IDTs.

The overall consistency between Fig. 5b and Fig. 4a justifies our AC lock-in measurement scheme of alternating SAW propagation in Fig. 5a.

The second harmonic acoustoelectric voltage $V_{2\mathrm{F}}$ probes the difference between two oppositely traveling SAW signals. Comparing $2V_{2\mathrm{F}}$ to $V_{1\mathrm{F}}$ in Fig. 5b (Methods), both $V_{1\mathrm{F}}$ and $V_{2\mathrm{F}}$ have spectral weights at 480 and 485 MHz, but of different proportions. As illustrated in Fig. 5b, $2V_{2\mathrm{F}}$ has an identical amplitude to $V_{1\mathrm{F}}$ at 480 MHz, which means one of the SAW states is effectively zero as indicated in Fig. 4a. In experiments, there could exist physical differences between two IDTs because of variations during the microfabrication process. This can cause differences in converting RF power to sound waves, leading to non-zero $V_{2\mathrm{F}}$ based on trivial technicality. For that, we vary the power injected into the IDT for the $-y$ propagation, while keep the power injected into the IDT for the $+y$ propagation constant (Fig. 5c). While the 485 MHz feature of $V_{2\mathrm{F}}$ demonstrates an input power dependence, the 480 MHz feature of $V_{2\mathrm{F}}$ remains constant. This clearly indicates that the 480 MHz spectral feature is solely a SAW propagating towards the $-y$ direction. Thus, the second-harmonic AC-lockin technique convincingly captures the intrinsic non-reciprocal SAWs and associated acoustoelectric effect in the DC/audio-frequency limit.

**Discussion**

As Stroh's ℵ-matrix equation of motion is based on the generalized Hooke's law of linear response, including the piezoelectric effects [19, 24, 25], it does not take the reflection effect into consideration. As ℵ is a real matrix, its eigenvalues appear in pairs of complex conjugates $p_\gamma = p_{\gamma+4}^*$ ($\gamma = 1,2,3,4$), and eigenstates are also complex conjugate to each other $\boldsymbol{\xi}_\gamma = \boldsymbol{\xi}_{\gamma+4}^*$, representing two oppositely propagating acoustic waves (Supplementary Materials). Only those eigenvalues $p_\alpha$ with an imaginary component that would lead to a surface wave decaying into the depth of bulk can construct the surface acoustic wave state by meeting the surface boundary conditions of both traction and the electric displacement field. The constructed SAW states by the ℵ-matrix equation would always be time reversal symmetric (Supplementary Materials).

For both $LiNbO_3$ and $LiTaO_3$, reciprocal and non-reciprocal SAW states have differently structured eigenvalues and eigenstates. When either $\hat{\boldsymbol{m}}$ or $\hat{\boldsymbol{n}}$ is perpendicular to a mirror plane of the bulk (such as along the $x$-axis), the ℵ matrix becomes traceless, forcing $p_\gamma = -p_{\gamma+4}$. So all eigenvalues are purely imaginary. In addition, each eigenstate $\boldsymbol{\xi}$ would have components with phase differences fixed at multiples of $\pi/2$ (Supplementary Materials). Conversely, when neither $\hat{\boldsymbol{m}}$ nor $\hat{\boldsymbol{n}}$ is perpendicular to a mirror plane, ℵ is no longer traceless and each eigenvalue can have both real and imaginary components. Each eigenstate would have phase differences not bound to multiples of $\pi/2$ between its components (Supplementary Materials). For those states, mechanical reflection can induce observable NSPUDT effects through the mechanical reflection coefficient $R_m^{(\pm)}$ [2].

The key symmetry element of scenario (1) is applicable to the macroscopic half-space. In substrates of point group *3m*, it is the mirror plane perpendicular to the substrate surface; for quartz in point group *32*, it is a two-fold rotational axis perpendicular to the substrate surface. In general, a rotational axis perpendicular to the substrate surface would remain as a good symmetry operation for the half-space substrate as it is for the bulk. As a surface acoustic wave that only depends on coordinates $x_1$ and $x_3$ along $\hat{\boldsymbol{m}}$ and $\hat{\boldsymbol{n}}$ respectively but not the transverse coordinate $x_2$ (Supplementary Materials), both symmetry operations of the mirror plane perpendicular to $\hat{\boldsymbol{m}}$ and an even-order rotational axis parallel to $\hat{\boldsymbol{n}}$ give the same transformation of $x_1 \to -x_1$ and $x_3 \to x_3$, while the change of $x_2$ is irrelevant to SAW states. Either of these symmetry elements can connect two oppositely propagating SAW states to preserve the reciprocity. In addition to the global symmetry element for scenario (1), Stroh's equation provides a symmetry protection to our scenario (2); as we will see below, that symmetry exists locally over nanoscale distances.

Ref. [19] approaches reciprocal SAW states from a different perspective of standing waves. As a resonance in the periodic IDT structure, a SAW state can have its frequency determined by either zeros or poles of the harmonic admittance under *open-* or *short-circuit* configurations respectively [19, 27, 28]. Ref. [19] argues that when frequency positions of a zero and a pole are degenerate, there would simultaneously be no current nor voltage on the electrodes, signaling the presence of a pure standing wave state with zero-valued nodes, which is made of oppositely traveling SAW states of equal amplitudes. Such a standing wave state is also the reciprocal SAW states. Although the admittance can be analytically expressed through eigenstates of the ℵ matrix, it was constructed in Ref. [19] by numerical solutions of the SAW state through coupling-of-modes equations including the multiple reflection process; the configurations for the standing wave were deduced from precision-limited null results. While the approach in Ref. [19] reveals previously unnoticed configurations in Ref. [18], it did not offer a clear understanding of the symmetry effects and the connection between two scenarios. We notice the determinant $\Delta^{(q)}$ in Eq. 22 of Ref. [19] would still have both zero and non-zero values for ℵ matrix eigenstates, which would all have the time reversal symmetry in the absence of the multiple reflection. This parameter $\Delta^{(q)}$ is thus not associated with standing wave states *in general* but rather represents an alternative statement about the two types of ℵ matrix eigenstates.

Beyond the ℵ matrix analysis, mechanical reflection in real IDT devices would reveal the presence of non-reciprocity as electrodes always have a finite thickness and mass. As the spectral width narrows, the non-reciprocal AE effect demonstrates a double-peak shape (Figs. 3-5), which contrasts the single-peak spectra of reciprocal AE effects. The difference in peak numbers reflects the pole-zero degeneracy/non-degeneracy discussed in Ref. [19]. The simple expression of $r^{(\pm)} = R_e^{(\pm)} + R_m^{(\pm)}\, h_{\mathrm{metal}}/\lambda$ treats electric and mass loading independently, as only the mechanical term contributes to the NSPUDT effect [2]. From our measurements in Figs. 3f, 4a-4d, and 5b, non-reciprocity behavior is also present at the high-frequency peak, which is regarded as under only the influence of electric reflection [2]. This emergent non-reciprocity at the high frequency peak is especially clear through the series of devices in Figs. 3b, 3d, and 3f, as $N$ remains constant while $h_{\mathrm{metal}}$ increases. When the mechanical reflection becomes strong, both reflection effects are no longer small perturbative corrections, so the simple approximation of $r^{(\pm)}$ breaks down.

Stroh's equation relies on two phenomena, the continuity of media $\partial\sigma_{ij}/\partial x_j = \rho\partial^2 u_i/\partial t^2$, and the Hooke's law $\sigma_{ij} = C_{ijkl}\,\partial u_k/\partial x_l$, where $\sigma_{ij}$ is the stress tensor, $u_i$ is the SAW displacement, and $x_i$ $(i = 1,2,3)$ the real space coordinates. They combine to form the equation of motion $C_{ijkl}\,\partial^2 u_k/\partial x_l \partial x_j = \rho\,\partial^2 u_i/\partial t^2$ as the basis of Stroh's 6×6 tensor, and can be further extended to the 8×8 tensor ℵ with piezoelectric and dielectric tensors. A salient feature of this equation of motion is that it has no intrinsic length scale, or rather that, while the speed of SAW is subsonic to the bulk speed of sound, it remains wavelength independent. On the other hand, the Hooke's law only requires properties such as the elastic tensor $C_{ijkl}$ to hold at the local scale, which can be much shorter than the wavelength.

The bulk symmetry property is regarded as preserved for the equation of motion. Only approaching the substrate surface, such a perspective of local bulk symmetry would have encountered a potential conflict, if not resolved by the boundary condition that requires both the traction $\boldsymbol{L}$ and electric displacement field component $D_n$ to approach zero at the surface.

The symmetry element in our scenario (1) can directly connect reciprocal SAW states, while the existing bulk symmetry elements in the configuration of scenario (2) could not. Instead, in the derivation of Stroh's matrix $\aleph$, it is the shear part of the strain tensor $s_{kl} = \partial u_k / \partial x_l + \partial u_l / \partial x_k$ that provides the symmetric tensor form [1] and in turn, sustains the equivalency between scenarios (1) and (2). Indeed, a SAW of Rayleigh type is necessarily made of a compressional motion along $\widehat{\boldsymbol{m}}$ and a shear motion along $\widehat{\boldsymbol{n}}$. As the compressional motion along $\widehat{\boldsymbol{m}}$ in scenario (1) becomes the shear motion along $\widehat{\boldsymbol{n}}$ in (2), and vice versa, an interchangeability exists between these two scenarios with a one-to-one correspondence. The symmetry element for scenario (1) holds at the global scale exceeding the wavelength distance, which matches with the perspective of hydrodynamic continuity. The symmetry in the strain tensor to ensure the second scenario only needs to hold locally at the nanometer scale. Through this perspective, the second scenario can be regarded as relying on a hidden, local symmetry.

## Methods:

**Microfabrication of SAW devices**. Our SAW devices to explore the acoustic-electronic effect are composed of two parts, an interdigital transducer that generates the propagating SAW states and a thin metal film to measure the AE effect; both are built on top of a piezoelectric substrate, of either $LiNbO_3$ or $LiTaO_3$. Substrates of different cuts (orientations) were purchased from several commercial vendors and growers. All SAW devices were fabricated in a cleanroom using maskless photolithography that achieves a minimum feature size of 2 $\mu$m. The fabrication process consisted of three lithographic stages, first the IDT, then the metal film, and finally electrode pads for wire bonding. The metal deposition was done using an electron-beam evaporation system. The IDTs are built with parameters listed in Figs. 1-5, first with a 5nm Ti adhesive layer, then either an Al or Au layer as the bulk of IDT, then a topping layer of 5nm Au for the protection of oxidation in case of Al is used for IDT. The thickness of the metal film should be about 3-7 nm [29], in order to generate a strong coupling to the AE effect. We chose the thin metal film to be made of non-magnetic elemental metals in a bi-layer structure of Ti (2 nm) and Au (3 nm). The electrode pads are made of Au of 100 nm thickness. To avoid edge effects of the device that reflects waves back towards the IDT, we placed small superglue droplets on the SAW pathway outside of the device region to provide effective damping and dissipation. Each piezoelectric substrate is fixed to a printed circuit board of planar RF waveguides by supergluing its corners, and they are electrically connected by 25 $\mu$m Au wire bonding. The printed circuit board uses surface-mount SMA connectors to further connect to RF power sources, and to VNA devices for the initial device characterization. The AE signals in the AC frequency range are picked up through normal pin connectors.

**AC lock-in measurements for the acoustoelectric effect.** Many groups measured the AE effect using a DC electrical current source. Here we measure the AE effect using the lock-in technique in the audio-frequency range in order to suppress the background noise and achieve a high sensitivity of signals below 10 nV (Fig. 4).

To establish whether there exists non-reciprocity, we employ a single, symmetric IDT measurement scheme outlined in Fig. 1a, featuring the IDT as the identical source for two oppositely propagating SAWs. This avoids microfabrication differences between two IDTs. In the measurement scheme of Fig. 1a, an arbitrary waveform generator (SDG2082X, Siglent

Technologies Co.) provides an AC frequency time series (~49 Hz) that has the amplitude of the modulus of a sine wave, which in turn amplitude modulates a RF signal generator (SG386, Stanford Research Systems) for the excitation of the IDT. We typically use a +7dBm input RF power for the SAW excitation, while the signal generator has a maximum power of +7dBm (~5mW). Given the weak AE signal in the presence of ample RF noises, a home-built low-pass filter using commercial elements with a cut-off frequency of 500 kHz (ELKEA103FA, Panasonic Inc.) was first applied to suppress the RF noise and prevent preamplifier saturation. The low pass filtered signal is further amplified by a low noise pre-amplifier (SRS560, Stanford Research Systems) with a typical amplification factor of 200×. Both the pre-amplified AE signal and the reference sine signal from the original arbitrary waveform generator are passed on to the lock-in amplifier (SR830, Stanford Research Systems), which extracts the AE voltage from the second harmonic channel of the reference sine wave.

This AC-frequency lock-in approach can be applied to directly measure the non-reciprocal AE signal (schematics outlined in Fig. 5a). Here two IDTs are utilized to provide alternatively and oppositely propagating SAWs towards the conductive pad in the center. The waveform evolution at each stage is detailly outlined in Fig. 5a. As each IDT is driven by RF signals with a modulated amplitude of a half sine wave, the AE effect that is proportional to the power (instead of the amplitude) of RF signal, should have the voltage response made of a series of odd harmonics of the primary AC frequency, but not the even harmonics. On the other hand, if there exists non-reciprocal behavior, then the even harmonic components would show up with measurable intensities. As Fig. 5a explains, to make a fair comparison, one would compare $V_{1\mathrm{F}}$ with $2V_{2\mathrm{F}}$, as $V_{1\mathrm{F}} = (V_{\mathrm{L}} + V_{\mathrm{R}})/2$ and $2V_{2\mathrm{F}} = |V_{\mathrm{L}} - V_{\mathrm{R}}|/2$, where $V_{\mathrm{L}}$ and $V_{\mathrm{R}}$ are absolute values of the AE amplitudes from left and right IDTs respectively.

The scheme in Fig. 5a also takes into consideration of practical complications. While we often consider the IDTs are perfect copies of each other, in the reality, they can have finite differences during micro-fabrication. So in our scheme, each IDT is driven by an independent RF power source, so that we can adjust the power input balance between two directions. A varying power study is exemplified in Fig. 5b and can serve as a method to demonstrate both the complete and incomplete NSPUDT effects at ~480 MHz and ~485 MHz respectively.

**Figure captions:**

**Fig. 1. AC lock-in measurement of the acoustoelectric effect.** (a) A schematic of the electrical circuitry. The RF excitation is AC frequency modulated, before injected into the SAW device. A representative device image is presented together with a zoomed-in view of IDT. The two SAWs propagate along opposite directions as marked by blue arrows. Two metal film pads are placed at equal distances from the IDT as probes of the acoustoelectric voltage, which is filtered and amplified before reaching the lock-in amplifier (Methods). (b) Schematics of the device geometry, which is defined by the surface normal $\hat{\boldsymbol{n}}$ and SAW propagation direction $\hat{\boldsymbol{m}}$. In general, SAW simultaneously propagate along both $\hat{\boldsymbol{m}}$ and $-\hat{\boldsymbol{m}}$. $\hat{\boldsymbol{n}}$ and $\hat{\boldsymbol{m}}$ are uniquely specified through Euler angles as listed in each data figure. (c) Acoustoelectric voltage measured at one metallic pad as a function of the input radiofrequency, for two devices at both low and high limits of IDT pair number $N$. The black curves are fits to the data, using a Sinc-squared function for the $N$ =20 device, and a Lorentzian function for the $N$ =250 device. The fitting of the Sinc-squared form is carried out using the sidebands (solid points), as the main spectral peak suffers a suppression from IDT mass loading.

**Fig. 2. Reciprocal acoustoelectric effect.** Measured AE effect from SAW states propagating along $\pm x$-axes of 128YX-cut $LiNbO_3$. Each panel of (a-d) corresponds to a different device with systematically adjusted IDT parameters $N$ and $h_{\mathrm{metal}}$. There is no difference in the AE effect between the two opposite directions, corresponding to the reciprocal relationship.

**Fig. 3**. **Non-reciprocal acoustoelectric effect.** Measured AE effect from SAW states propagating along $\pm Y$-axes of 128YX-cut $LiNbO_3$. Each panel of (a-f) corresponds to a different device with varying IDT parameters $N$ and $h_{\mathrm{metal}}$. Here, the red and blue curves corresponding to SAW states propagating in opposite directions, demonstrate the non-reciprocity, as the difference between them grows with either increasing $N$ or $h_{\mathrm{metal}}$.

**Fig. 4. Reciprocity based on choices of the substrate and propagation direction.** Eight devices with different choices of substrate, surface normal, and propagation direction are tested. Both large IDT $N$ and $h_{\mathrm{metal}}$ are employed for each device to amplify the potential non-reciprocity of the oppositely propagating SAWs. Panels (a-d) demonstrate non-reciprocal AE effect, while panels (e-h) demonstrate reciprocal scenarios. All our configuration/substrate choices would not cause beam steering to the propagating SAW states [22].

**Fig. 5. Direct measurement of the non-reciprocal AE effect with the second harmonic AC probe.** (a) A schematic of the circuitry to isolate the non-reciprocal AE effect. With a two-IDT device, alternative RF signals are sent into the device from two sides. The alternating AE potential generated at the detecting pad is analyzed by the AC lock-in technique (Methods). The primary frequency signal $V_{1\mathrm{F}}$ and second harmonic signal $V_{2\mathrm{F}}$ are separately extracted. Schematics of the wave form demonstrate the signal characteristics; $V_{1\mathrm{F}}$ should be compared to $2V_{2\mathrm{F}}$ based on the measurement scheme (Methods). (b) A comparison of measured $V_{1\mathrm{F}}$ and $V_{2\mathrm{F}}$ from a device of identical condition to that of Fig. 4a, using equal input power of 5dBm on two sides. Different spectral weight ratios between $V_{1\mathrm{F}}$ and $V_{2\mathrm{F}}$ indicate two regions of interest at 480 and 485 MHz. (c) With input RF powers fixed along the $-y$ direction and varied along the $+y$ direction, the spectral weight of $V_{2\mathrm{F}}$ remains constant at 480 MHz but varies around a non-zero value at 485 MHz. So SAW propagating along the $+y$ direction has no spectral weight in the 480 MHz region, which hosts a complete NSPUDT effect.

**Acknowledgement:** We are grateful to insightful discussions with Woowon Kang. We acknowledge financial support from the Okinawa Institute of Science and Technology Graduate University (OIST), with subsidy funding from the Cabinet Office, Government of Japan. We also thank the Mechanical Engineering and Microfabrication Support Section of OIST for the usage of shared cleanroom facility.

**Author Contributions:** Y. F. designed research. All authors performed research. S.V. and Y. F. analyzed data and prepared the manuscript. All authors commented.

**Competing Interest Statement:** The authors declare no competing interests.

**Data availability:** All study data are included in the article and/or supplementary materials.

**References:**

[1] S. Datta, *Surface Acoustic Wave Devices*, Prentice Hall, London (1986).

[2] D. Morgan, *Surface Acoustic Wave Filters*, Second edition, Academic Press, Oxford (2007).

[3] R.L. Willett, M.A. Paalanen, R.R. Ruel, K.W. West, L.N. Pfeiffer, D.J. Bishop, Anomalous sound propagation at ν =1/2 in a 2D electron gas: observation of a spontaneously broken translational symmetry? *Phys. Rev. Lett*. **65**, 112-115 (1990).

[4] A. Bienfait, Y.P. Zhong, H.-S. Chang, M.-H. Chou, C.R. Conner, É. Dumur, J. Grebel, G.A. Peairs, R.G. Povey, K.J. Satzinger, A.N. Cleland, Quantum erasure using entangled acoustic phonons. *Phys. Rev. X* **10**, 021055 (2020).

[5] R.P.G. McNeil, M. Kataoka, C.J.B. Ford, C.H.W. Barnes, D. Anderson, G.A.C. Jones, I. Farrer, D.A. Ritchie, On-demand single-electron transfer between distant quantum dots, *Nature* **477**, 439-442 (2011).

[6] A. Huang, H. Liu, O. Manor, P. Liu, J. Friend, Enabling rapid charging lithium metal batteries via surface acoustic wave-driven electrolyte flow. *Adv. Mater.* **32**, 1907516 (2020).

[7] R. Chen, C. Chen, L. Han, P. Liu, R. Su, W. Zhu, Y. Zhou, F. Pan, C. Song. Ordered creation and motion of skyrmions with surface acoustic wave. *Nat. Commun.* **14**, 4427 (2023).

[8] P. Zhang, H. Bachman, A. Ozcelik, T.J. Huang, Acoustic microfluids, *Annu. Rev. Anal. Chem*. **13**, 17-43 (2020).

[9] R. H. Parmenter, Acoustoelectric effect. *Phys. Rev*. **89**, 990-998 (1953).

[10] G. Weinreich, T.M. Sanders, H.G. White, Acoustoelectric effect in *n*-type germanium. *Phys. Rev.* **114,** 33-44 (1959).

[11] T. Kawada, M. Kawaguchi, M. Hayashi, Unidirectional planar Hall voltages induced by surface acoustic waves in ferromagnetic thin films. *Phys. Rev. B* **99**, 184435 (2019).

[12] D. Castilla, M. Muñoz, M. Sinusía, R. Yanes, J.L. Prieto, Large asymmetry in the magnetoresistance loops of ferromagnetic nanostrips induced by Surface Acoustic Waves. *Sci. Rep.* **11**, 8586 (2021).

[13] L. Liao, F. Chen, J. Puebla, J. Kishine, K. Kondou, W. Lou, D. Zhao, Y. Zhang, Y. Ba, Y. Otani, Nonreciprocal magnetoacoustic waves with out-of-plane phononic angular momenta. *Sci. Adv.* **10**, eado2504 (2024).

[14] C. Chen, L. Han, P. Liu, Y. Zhang, S. Liang, Y. Zhou, W. Zhu, S. Fu, F. Pan, C. Song, Direct-current electrical detection of surface-acoustic-wave-driven ferromagnetic resonance. *Adv. Mater*. 35, 2302454 (2023).

[15] Y. Ilisavskii, A. Goltsev, K. Dyakonov, V. Popov, E. Yakhkind, V.P. Dyakonov, P. Gierłowski, A. Klimov, S.J. Lewandowski, H. Szymczak. Anomalous acoustoelectric effect in $La_{0.67}Ca_{0.33}MnO_3$ films. *Phys. Rev. Lett.* **87**, 146602 (2001).

[16] E. Preciado, F. Schülein, A. Nguyen, D. Barroso, M. Isarraraz, G. von Son, I. Lu, W. Michailow, B. Möller, V. Klee, J. Mann, A. Wixforth, L. Bartels, H.J. Krenner, Scalable fabrication of a hybrid field-effect and acousto-electric device by direct growth of monolayer $MoS_2/LiNbO_3$. *Nat. Commun.* **6**, 8593 (2015).

[17] P.V. Wright, The natural single-phase unidirectional transducer: a new low-loss SAW transducer. *1985 IEEE Ultrasonics Symposium,* 58-63 (1985).

[18] G. Martin, M. Weihnacht, Properties of interdigital transducers in relation to the substrate crystal symmetry. *IEEE Trans. Ultrasonics, Ferroelectrics, and Frequency Control* **43**, 646–653 (1996).

[19] A.N. Darinskii, S.V. Biryukov, M. Weihnacht, Fundamental frequency degeneracy of standing surface acoustic waves under metallic gratings on piezoelectric substrates. *J. Acoust. Soc. Am.* **112**, 2003-2013 (2002).

[20] G. Martin, H. Schmidt, B. Wall, Natural SPUDT behavior of YZ $LiNbO_3$. *2010 IEEE International Ultrasonics Symposium*, 189-192 (2010).

[21] V.P. Plessky, SAW impedance elements. *IEEE Trans. Ultrason. Ferroelectr. Freq. Control* **42**, 870-875 (1995).

[22] J.A.J. Slobodnik, *The temperature coefficients of acoustic surface wave velocity and delay on lithium niobate, lithium tantalate, quartz, and tellurium dioxide*. Air Force Cambridge Research Laboratories, Office of Aerospace Research, United States Air Force (1971).

[23] V. Soluch, M. Lysakowska, Surface acoustic waves on X-cut $LiNbO_3$. *IEEE Trans. Ultrasonics, Ferroelectrics, and Frequency Control* **52**, 145-147 (2005).

[24] A. N. Stroh, Steady state problems in anisotropic elasticity, *J. Math. Phys.* **41**, 77-103 (1962).

[25] J. Lothe, D.M. Barnett, Integral formalism for surface waves in piezoelectric crystals. Existence considerations, *J. Appl. Phys.* **47**, 1799-1807 (1976).

[26] N. Zhang, J. Mei, T. Gopesh, J. Friend, Optimized, omnidirectional surface acoustic wave source: 152°*Y*-rotated cut of lithium niobate for acoustofluidics, *IEEE Trans. Ultrasonics, Ferroelectrics, and Frequency Control* **67**, 2176-2186 (2020).

[27] Y. Zhang, J. Desbois, L. Boyer, Characteristic parameters of surface acoustic waves in a periodic metal grating on a piezoelectric substrate. *IEEE Trans. Ultrasonics, Ferroelectrics, and Frequency Control* **40**, 183-192 (1993).

[28] V.P. Plessky, S.V. Biryukov, J. Kosela, Harmonic admittance and dispersion equations – the theorem. *IEEE Trans. Ultrasonics, Ferroelectrics, and Frequency Control* **49**, 528-534 (2002).

[29] P. Bierbaum, Interaction of ultrasonic surface waves with conduction electrons in thin metal films, *Appl. Phys. Lett.* **21**, 595-598 (1972).

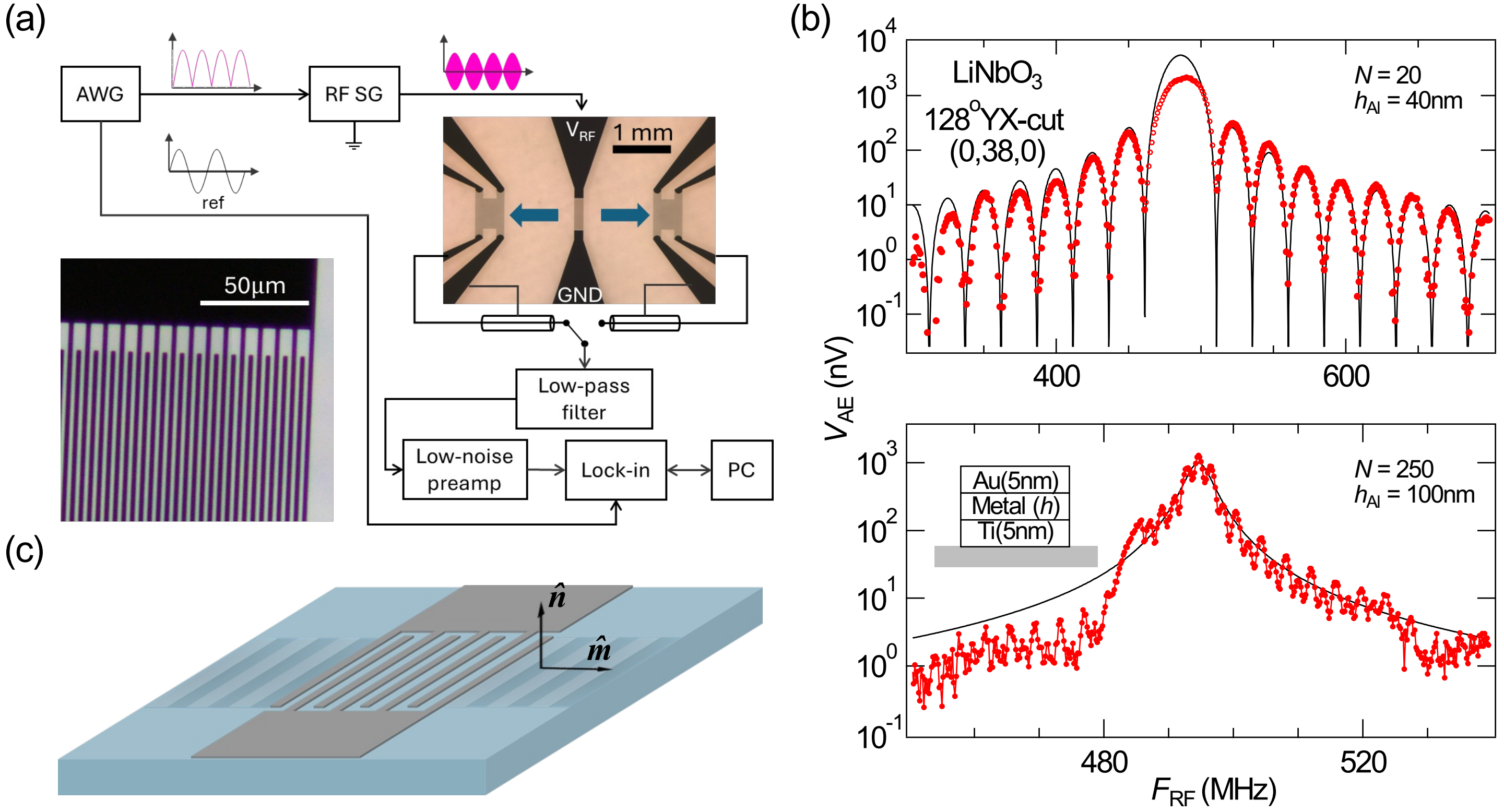

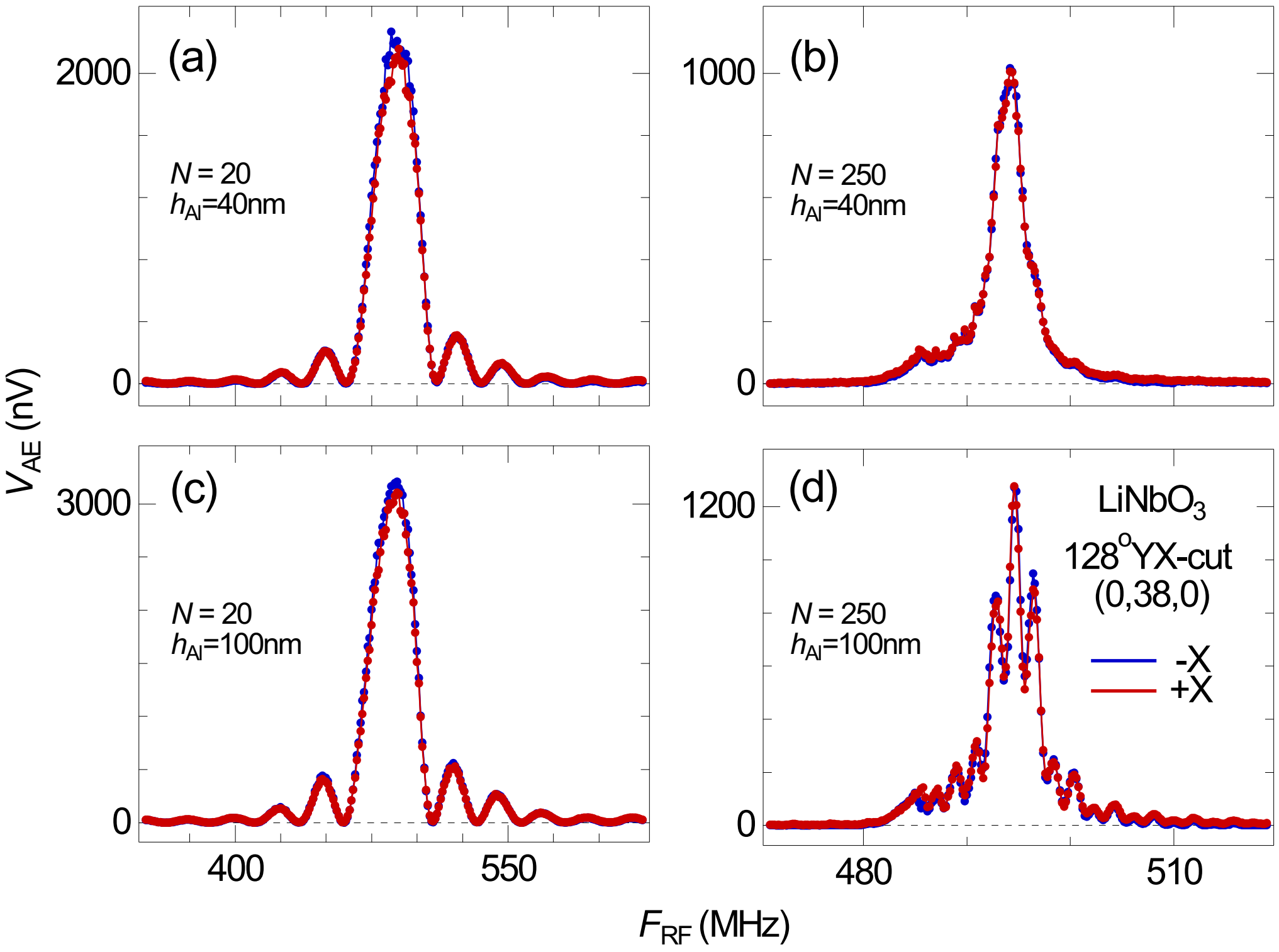

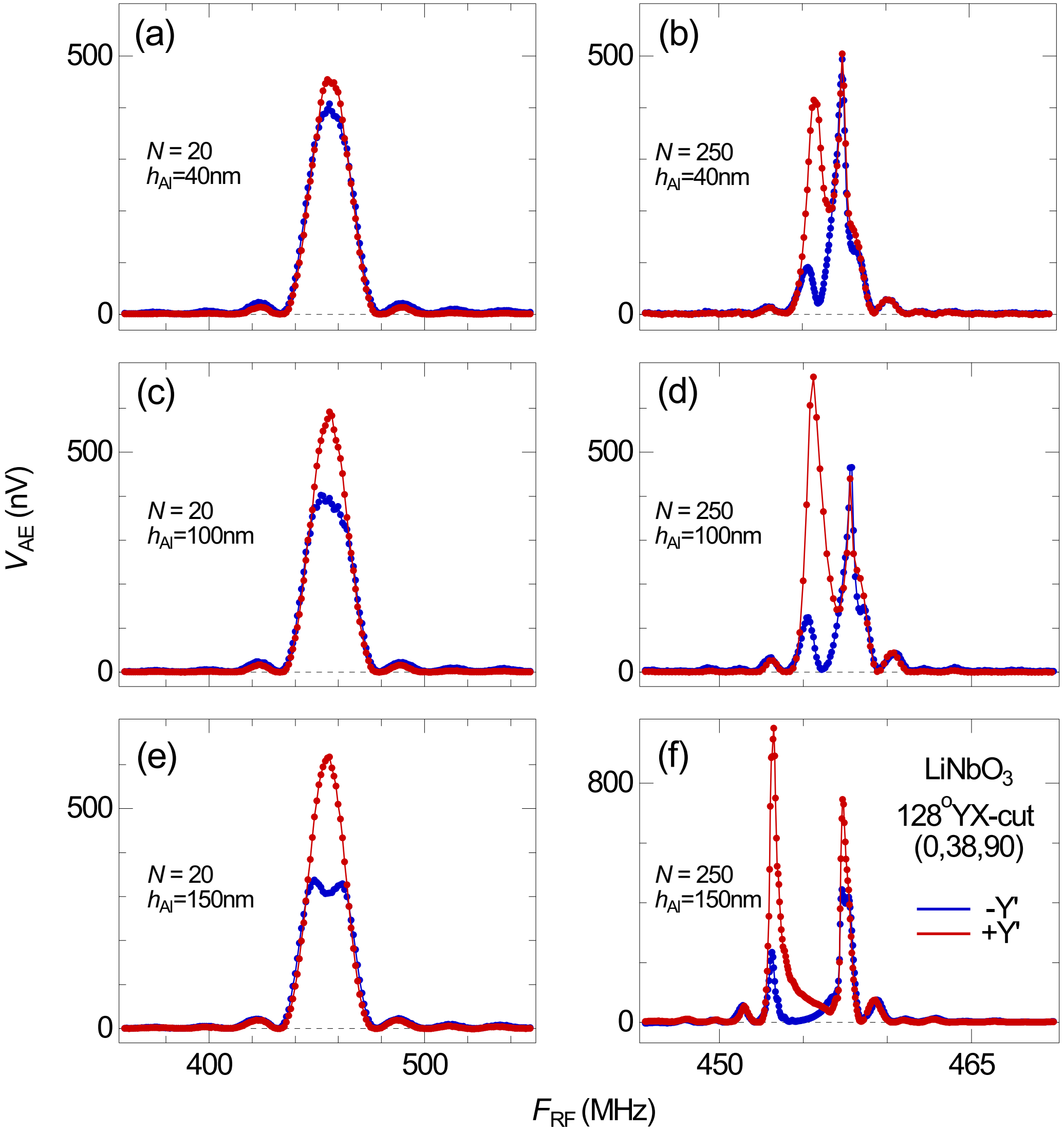

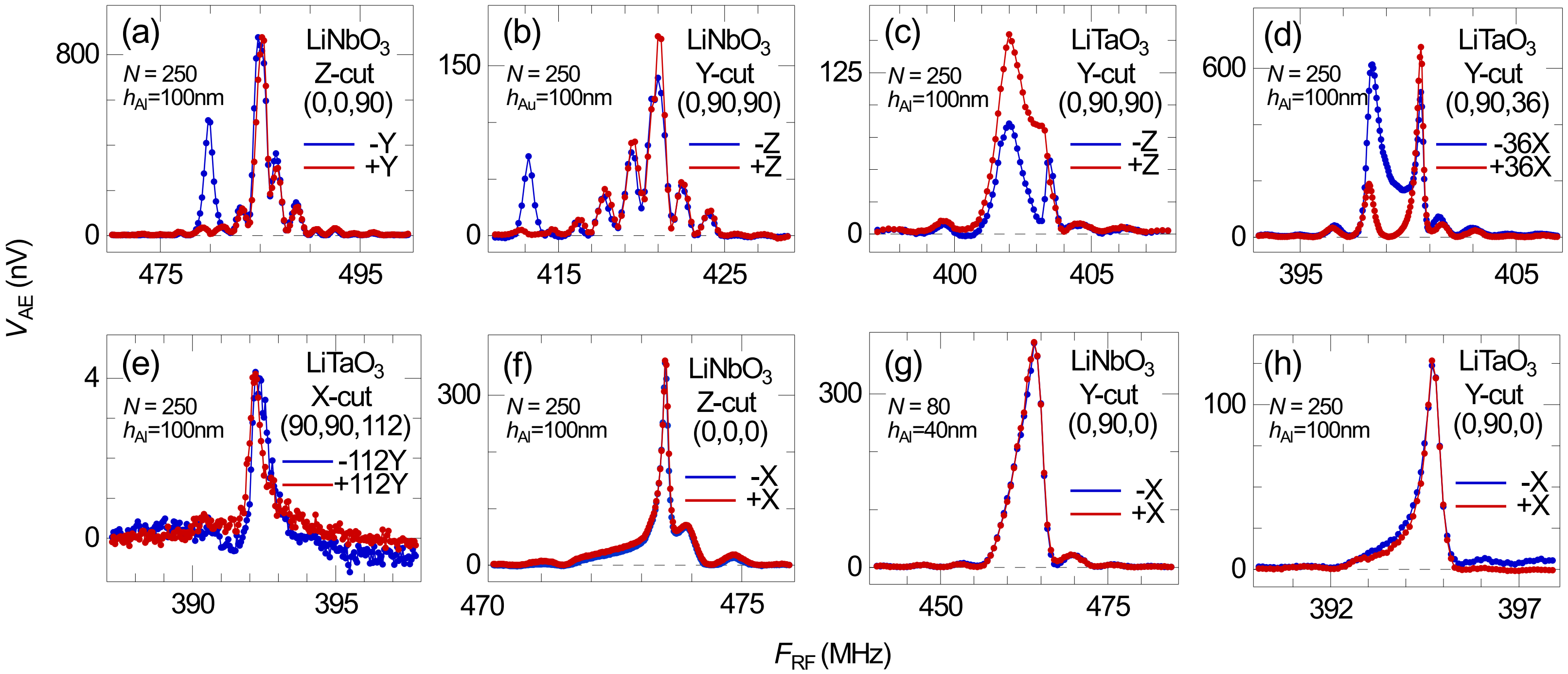

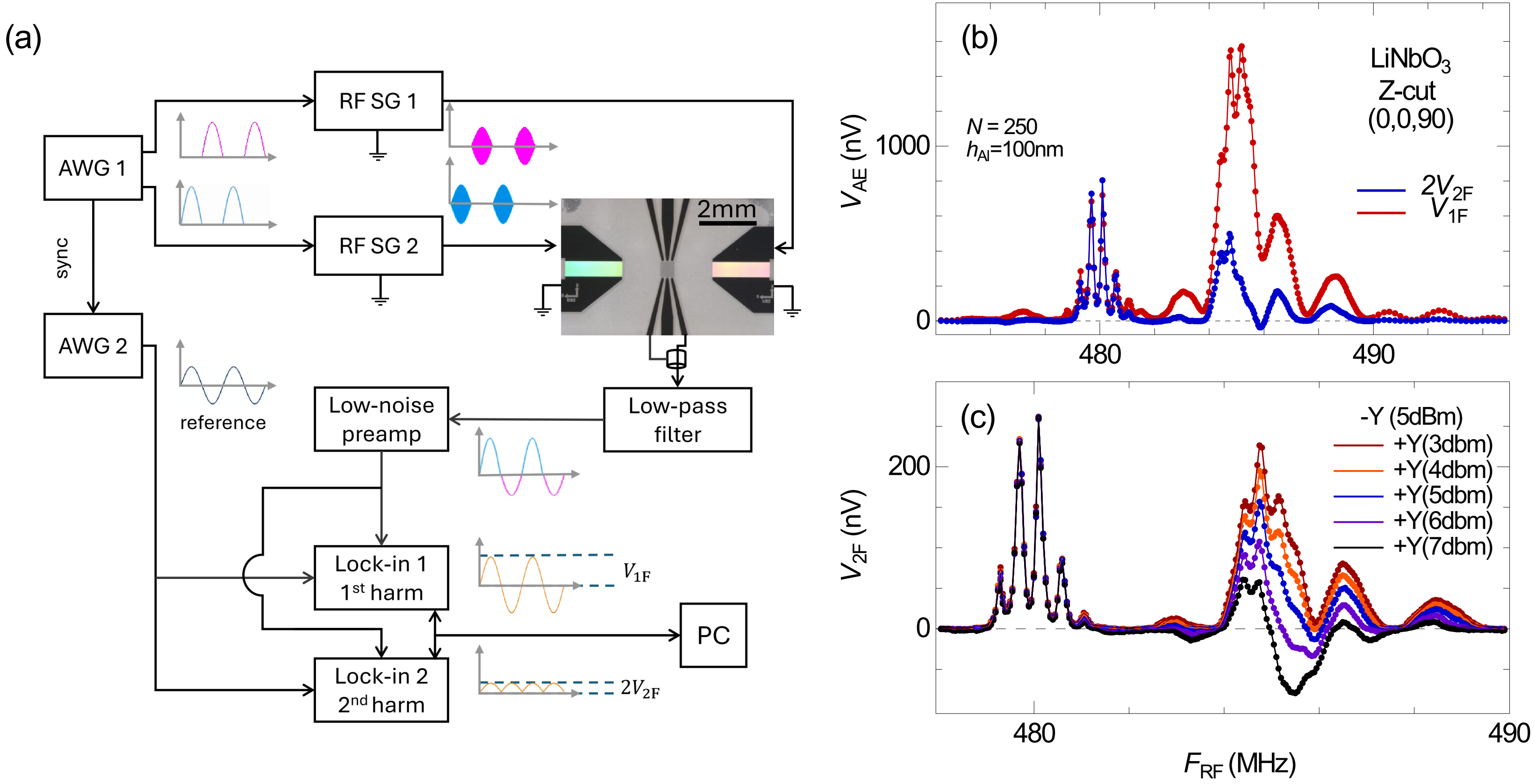

**Supplementary Materials for:**

**Nanoscale symmetry protection of the reciprocal acoustoelectric effect**

Sandeep Vijayan[1], Stephan Suffit[1], Scott E. Cooper[1], Yejun Feng[1,*]

[1]Okinawa Institute of Science and Technology Graduate University, Onna, Okinawa 904-0495, Japan

*Corresponding author: yejun@oist.jp

**Supplementary note: Surface acoustic waves in a substrate of point group *3m***

## 1. SAW states as eigenstate solutions of a matrix formalism

In 1960's, Stroh [24] provided a formalism to represent the elasto-dynamic equations of an anisotropic solid as an eigenvalue problem of a 6×6 matrix, known as the Stroh matrix. The matrix only depends on the elasticity tensor, the velocity of the wave in the medium, and the density of the medium. This formalism emerges from a pair of vector equations that couple two three-dimensional vectors, namely the displacement vector and the traction vector. With the presence of both vectors, a six-dimensional state vector can be formed, which serves as the eigenvector of the Stroh matrix. Lothe and Barnett [25] further improved the formalism by incorporating the electrostatic equations along with the elasto-dynamic equations, assuming a semi-infinite piezoelectric solid. This addition morphs the $6 \times 6$ Stroh matrix into an $8 \times 8$ matrix, which also requires the extra knowledge of the piezoelectric tensor and the permittivity tensor for its formation. Each eight-dimensional eigenvector is composed of the displacement vector $\boldsymbol{A}$, electrostatic potential $\phi$, traction vector $\boldsymbol{L}$, and the normal component of the displacement field $D_n$.

Here we summarize the general formalism. We assume that a semi-infinite piezoelectric medium has a surface normal along the unit vector $\widehat{\boldsymbol{n}}$ and the wave is propagating along the unit vector $\widehat{\boldsymbol{m}}$ within the surface (Fig. 1). So, the SAW state can be treated as an eigenstate problem as:

$$\widehat{\boldsymbol{N}}\boldsymbol{\xi}_\alpha = p_\alpha \boldsymbol{\xi}_\alpha, \qquad \alpha = 1,2,\dots,8. \tag{S1}$$

Here, $p_\alpha$ are eigenvalues, $\boldsymbol{\xi}_\alpha$ are eigenfunctions, and $\widehat{\boldsymbol{N}}$ is an 8×8 matrix defined by four blocks of 4×4 matrices as:

$$\widehat{\boldsymbol{N}} = -\begin{bmatrix} (nn)^{-1}(nm) & (nn)^{-1} \\ (mn)(nn)^{-1}(nm) - (mm) & (mn)(nn)^{-1} \end{bmatrix}, \tag{S2}$$

with the matrices $(nn), (mn), (nm)$ and $(mm)$ defined in the following manner [24],

$$(ab) = \begin{bmatrix} a_k c_{k11l} b_l - \rho v^2 a_k m_k m_l b_l & a_k c_{k12l} b_l & a_k c_{k13l} b_l & a_k e_{l1k} b_l \\ a_k c_{k21l} b_l & a_k c_{k22l} b_l - \rho v^2 a_k m_k m_l b_l & a_k c_{k23l} b_l & a_k e_{l2k} b_l \\ a_k c_{k31l} b_l & a_k c_{k32l} b_l & a_k c_{k33l} b_l - \rho v^2 a_k m_k m_l b_l & a_k e_{l3k} b_l \\ a_k e_{k1l} b_l & a_k e_{k2l} b_l & a_k e_{k3l} b_l & -a_k \varepsilon_{kl} b_l \end{bmatrix}.$$

Here, $\boldsymbol{c}, \boldsymbol{e}$ and $\boldsymbol{\varepsilon}$ are the elasticity, piezoelectric, and permittivity tensors respectively; $v$ is the phase velocity of the surface acoustic wave along $\widehat{\boldsymbol{m}} = (m_1, m_2, m_3)$, $\rho$ is the mass density of the material, and the summation runs over indices $k$ and $l$ from 1 to 3. At this stage, the coordination system (1, 2, 3) represents the three-dimensional space and can be rather general as long as the three axes are orthogonal. Both $\widehat{\boldsymbol{m}}$ and $\widehat{\boldsymbol{n}}$ vectors are projected onto this coordinate system.

From Equation S2, we define,

$$\hat{\boldsymbol{A}} = -(nn)^{-1}(nm),$$
$$\hat{\boldsymbol{B}} = -(nn)^{-1},$$
$$\hat{\boldsymbol{C}} = -(mn)(nn)^{-1}(nm) + (mm). \tag{S3}$$

Then Equation S2 can be expressed as,

$$\hat{\boldsymbol{N}} = \begin{bmatrix} \hat{\boldsymbol{A}} & \hat{\boldsymbol{B}} \\ \hat{\boldsymbol{C}} & \hat{\boldsymbol{A}}^{\boldsymbol{T}} \end{bmatrix}. \tag{S4}$$

The eigenvalue problem becomes:

$$\det(\hat{\boldsymbol{N}} - p_\alpha \hat{\boldsymbol{I}}_8) = 0. \tag{S5}$$

Using Schur's complement, we can rewrite Equation S5 as,

$$\det(\hat{\boldsymbol{A}} - p_\alpha \hat{\boldsymbol{I}}_4)\det\left(\hat{\boldsymbol{A}}^{\boldsymbol{T}} - p_\alpha \hat{\boldsymbol{I}}_4 - \hat{\boldsymbol{C}}(\hat{\boldsymbol{A}} - p_\alpha \hat{\boldsymbol{I}}_4)^{-1}\hat{\boldsymbol{B}}\right) = 0.$$

Since $p_\alpha$ is the eigenvalue for $\hat{\boldsymbol{N}}$ but not $\hat{\boldsymbol{A}}$, $\det(\hat{\boldsymbol{A}} - p_\alpha \hat{\boldsymbol{I}}_4) \neq 0$. We should have,

$$\det\left(\hat{\boldsymbol{A}}^{\boldsymbol{T}} - p_\alpha \hat{\boldsymbol{I}}_4 - \hat{\boldsymbol{C}}(\hat{\boldsymbol{A}} - p_\alpha \hat{\boldsymbol{I}}_4)^{-1}\hat{\boldsymbol{B}}\right) = 0,$$
$$\Rightarrow \det\left(\hat{\boldsymbol{A}}^{\boldsymbol{T}}\hat{\boldsymbol{B}}^{-\boldsymbol{1}} - p_\alpha \hat{\boldsymbol{B}}^{-\boldsymbol{1}} - \hat{\boldsymbol{C}}(\hat{\boldsymbol{A}} - p_\alpha \hat{\boldsymbol{I}}_4)^{-1}\right)\det(\hat{\boldsymbol{B}}) = 0, \qquad \because \det(\hat{\boldsymbol{B}}) \neq 0$$
$$\Rightarrow \det\left(\hat{\boldsymbol{A}}^{\boldsymbol{T}}\hat{\boldsymbol{B}}^{-\boldsymbol{1}}(\hat{\boldsymbol{A}} - p_\alpha \hat{\boldsymbol{I}}_4) - p_\alpha \hat{\boldsymbol{B}}^{-\boldsymbol{1}}(\hat{\boldsymbol{A}} - p_\alpha \hat{\boldsymbol{I}}_4) - \hat{\boldsymbol{C}}\right)\det\left((\hat{\boldsymbol{A}} - p_\alpha \hat{\boldsymbol{I}}_4)^{-1}\right) = 0,$$
$$\Rightarrow \det\left(p_\alpha^2 \hat{\boldsymbol{B}}^{-\boldsymbol{1}} - p_\alpha\left(\hat{\boldsymbol{A}}^{\boldsymbol{T}}\hat{\boldsymbol{B}}^{-\boldsymbol{1}} - \hat{\boldsymbol{B}}^{-\boldsymbol{1}}\hat{\boldsymbol{A}}\right) + \hat{\boldsymbol{A}}^{T}\hat{\boldsymbol{B}}^{-\boldsymbol{1}}\hat{\boldsymbol{A}} - \hat{\boldsymbol{C}}\right) = 0. \tag{S6}$$

Substitute Equations S3 into Equation S6, we get an eigenvalue equation:

$$\det(p_\alpha^2(nn) - p_\alpha((mn) + (nm)) + (mm)) = 0. \tag{S7}$$

With the coefficient matrices all real and symmetric, it is necessary that eigenvalues must be either real or occur in pairs of complex conjugates $(p_\gamma, p_\gamma^*)$, with $\gamma = 1,2,3,4$, and the eigenvectors $(\boldsymbol{\xi}_\gamma, \boldsymbol{\xi}_\gamma^*)$ must also occur in complex conjugate pairs.

As an 8×8 matrix, $\hat{\boldsymbol{N}}$ has eight eigenvalues $(p_\alpha)$ and eigenvector $(\boldsymbol{\xi}_\alpha)$, so there are eight modes, defined as:

$$\boldsymbol{\xi}_\alpha(\boldsymbol{r}, t) = \boldsymbol{\xi}_\alpha \exp[i\boldsymbol{k} \cdot (\hat{\boldsymbol{m}} r_1 + p_\alpha \hat{\boldsymbol{n}} r_3 - \boldsymbol{v}t)], \quad \alpha = 1,2,\ldots,8. \tag{S8}$$

Here $\hat{\boldsymbol{m}}$ and $\hat{\boldsymbol{n}}$ define a wave-based coordinate system ($\hat{\boldsymbol{m}}$, $\hat{\boldsymbol{m}} \times \hat{\boldsymbol{n}}$, $-\hat{\boldsymbol{n}}$ ) for the surface wave states (Fig. 1). The spatial position $\boldsymbol{r} = (r_1, r_2, r_3)$ is defined according.

All eight modes can be divided into two equal sized groups, with eigenvalues of one group always the complex conjugate of the other. The grouping scheme based on complex conjugacy separates waves going to one side ($\boldsymbol{k}$ and $\boldsymbol{v}$) and modes going to the opposite direction ($\boldsymbol{k}' = -\boldsymbol{k}$ and $\boldsymbol{v}' = -\boldsymbol{v}$). For a pair of oppositely propagating waves, we have $\boldsymbol{\xi}_\gamma(\boldsymbol{r}, t) = \boldsymbol{\xi}_\gamma \exp\left[i\boldsymbol{k} \cdot \left(\hat{\boldsymbol{m}} \cdot r_1 + p_\gamma \hat{\boldsymbol{n}} \cdot r_3 - \boldsymbol{v}t\right)\right]$, and

$$\boldsymbol{\xi}_{\gamma+4}(\boldsymbol{r}, t) = \boldsymbol{\xi}_{\gamma+4} \exp\left[-i\boldsymbol{k} \cdot \left(\hat{\boldsymbol{m}} \cdot r_1 + p_\gamma^* \hat{\boldsymbol{n}} \cdot r_3 + \boldsymbol{v}t\right)\right], \quad \gamma = 1,2,3,4.$$

Here we already used $p_{\gamma+4} = p_\gamma^*$, $\boldsymbol{k}' = -\boldsymbol{k}$, and $\boldsymbol{v}' = -\boldsymbol{v}$, and for the amplitude, we have $\boldsymbol{\xi}_\gamma = \boldsymbol{\xi}_{\gamma+4}^*$. The time reversal operation $\mathbf{T}$ transforms a state as $\mathbf{T}\boldsymbol{\xi}_\gamma(\boldsymbol{r}, t) = \boldsymbol{\xi}_\gamma^*(\boldsymbol{r}, -t)$. It is straightforward to verify that $\boldsymbol{\xi}_\gamma^*(\boldsymbol{r}, -t) = \boldsymbol{\xi}_{\gamma+4}(\boldsymbol{r}, t)$. So, the oppositely propagating $\boldsymbol{\xi}_\gamma(\boldsymbol{r}, t)$ and $\boldsymbol{\xi}_{\gamma+4}(\boldsymbol{r}, t)$ are always equivalent under the time reversal operation. This conforms to the time reversal symmetry for a charge system that one propagating mode is simply the time reversed mode of the oppositely propagating one.

Since our analysis is for SAWs, it becomes sufficient to focus on only the modes that correspond to truly complex eigenvalues. This is because those modes with real eigenvalues cannot decay into the bulk, and if included, each will result in a bulk wave.

The normalization of the eigenvectors takes the following appraoch. With the way that $\widehat{\boldsymbol{N}}$ is defined, we introduce the tensor $\widehat{\boldsymbol{T}}$ to make the product $\widehat{\boldsymbol{T}}\widehat{\boldsymbol{N}}$ symmetric, as,

$$\widehat{\boldsymbol{T}} = \begin{bmatrix} 0 & 0 & 0 & 0 & 1 & 0 & 0 & 0 \\ 0 & 0 & 0 & 0 & 0 & 1 & 0 & 0 \\ 0 & 0 & 0 & 0 & 0 & 0 & 1 & 0 \\ 0 & 0 & 0 & 0 & 0 & 0 & 0 & 1 \\ 1 & 0 & 0 & 0 & 0 & 0 & 0 & 0 \\ 0 & 1 & 0 & 0 & 0 & 0 & 0 & 0 \\ 0 & 0 & 1 & 0 & 0 & 0 & 0 & 0 \\ 0 & 0 & 0 & 1 & 0 & 0 & 0 & 0 \end{bmatrix}.$$

This allows us to rewrite equation S1 as,

$$\begin{aligned} &\widehat{\boldsymbol{T}}\widehat{\boldsymbol{N}}\boldsymbol{\xi}_\alpha = p_\alpha \widehat{\boldsymbol{T}}\boldsymbol{\xi}_\alpha \\ &\Rightarrow \boldsymbol{\xi}_\alpha^T \widehat{\boldsymbol{T}}\widehat{\boldsymbol{N}} = p_\alpha \boldsymbol{\xi}_\alpha^T \widehat{\boldsymbol{T}}. \end{aligned} \tag{S9}$$

This implies $\boldsymbol{\xi}_\alpha^T \widehat{\boldsymbol{T}}$ is the left eigenvector of $\widehat{\boldsymbol{N}}$, which allows for us to impose a self normalization condition on the eigenvectors as,

$$\boldsymbol{\xi}_\alpha^T \widehat{\boldsymbol{T}} \boldsymbol{\xi}_\beta = \delta_{\alpha\beta}\ , \qquad \alpha, \beta = 1,2,\ldots,8 \tag{S10}$$

This normalisation condition of eigenvectors leads to individual mode defined as,

$\boldsymbol{\xi}_\alpha = (A_{\alpha 1}, A_{\alpha 2}, A_{\alpha 3}, \phi_\alpha, L_{\alpha 1}, L_{\alpha 2}, L_{\alpha 3}, D_\alpha)^T$, where $A_{\alpha i}$ are components of the elastic displacement amplitude vector $\boldsymbol{A}_\alpha$, $\boldsymbol{L}_\alpha$ is proportional to the surface traction $\boldsymbol{f}_\alpha$ as $\boldsymbol{L}_\alpha = i\boldsymbol{f}_\alpha/k$, $\phi_\alpha$ is the electric potential and $D_\alpha$ is proportional to the normal component of the displacement field $D_{n\alpha}$ as $D_\alpha = iD_{n\alpha}/k$ [19, 25]. All are indexed consistently with a given eigenvalue $p_\alpha$.

From here on, we further analyze the $\widehat{\boldsymbol{N}}$ matrix for conditions leading to reciprocity and non-reciprocity (or symmetry vs. asymmetry) in SAW states, which reveals insights about the lattice symmetry.

## 2. Scenarios of reciprocal SAW states

In the current study, we have used substrates of $LiNbO_3$ and $LiTaO_3$, which are both of the same point group *3*m, and without the inversion symmetry. In this section, we examine all the substrate geometries that have either $\widehat{\boldsymbol{m}}$ or $\widehat{\boldsymbol{n}}$ vectors along the crystalline $x$-axis, which is perpendicular to the mirror plane as shown in Fig. S1. Instead of using the typical coordinate system ($\widehat{\boldsymbol{m}}$, $\widehat{\boldsymbol{m}} \times \widehat{\boldsymbol{n}}$, $-\widehat{\boldsymbol{n}}$ ), we express the $\widehat{\boldsymbol{N}}$ matrix in a sample-oriented coordinate system (1, 2, 3) = ($x, y', z'$). This coordinate system is designed to take the advantage that in this Supplementary Note, we only consider scenarios that the crystalline $x$-axis is parallel to one of the three orthogonal vectors $\widehat{\boldsymbol{m}}$, $\widehat{\boldsymbol{n}}$, (discussed here) or $\widehat{\boldsymbol{m}} \times \widehat{\boldsymbol{n}}$ (discussed in the next section). After the $x$-axis is matched, the remaining two orthogonal axes are within the crystalline $y$-$z$ plane and are related to the $y$ and $z$ axes by a rotational transformation. We thus define the remaining two orthogonal axes among $\widehat{\boldsymbol{m}}$, $\widehat{\boldsymbol{n}}$, and $\widehat{\boldsymbol{m}} \times \widehat{\boldsymbol{n}}$ as $y'$ and $z'$ to form the ($x$, $y'$, $z'$) coordinate system. These ($x, y', z'$) coordinates are essentially permutated ($\widehat{\boldsymbol{m}}$, $\widehat{\boldsymbol{m}} \times \widehat{\boldsymbol{n}}$, $-\widehat{\boldsymbol{n}}$ ) coordinates, but can better incorporate in the crystalline symmetry. For both cases considered in this Section, we have either $\widehat{\boldsymbol{m}}||x$ ($\widehat{\boldsymbol{n}}||z'$) or $\widehat{\boldsymbol{n}}||x$ ($\widehat{\boldsymbol{m}}||z'$). They both lead the $\widehat{\boldsymbol{N}}$ matrix to a general form as,

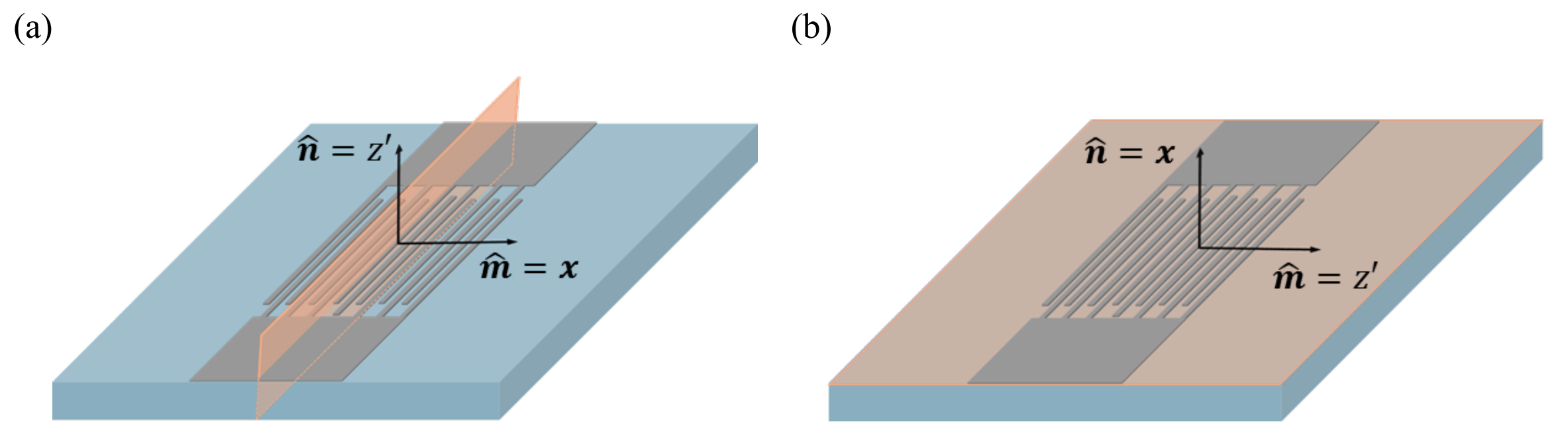

Figure S1: Schematics of reciprocal SAW states. (a) Propagation vector $\hat{\boldsymbol{m}}$ is along the crystalline $x$-axis, and an arbitrary direction $z'$ in the $y$-$z$ plane can be chosen as the substrate surface normal $\hat{\boldsymbol{n}}$, provided the beam steering effect is not a concern. (b) Conversely, the substrate surface normal $\hat{\boldsymbol{n}}$ is parallel to the crystalline $x$-axis and the propagation is along an arbitrary direction $z'$ within the $y$-$z$ plane. The orange plane represents the mirror $y$-$z$ plane of the bulk crystal.

$$\widehat{\boldsymbol{N}} = \begin{bmatrix} 0 & \times & \times & \times & \times & 0 & 0 & 0 \\ 0 & 0 & 0 & 0 & 0 & \times & \times & \times \\ 1 & 0 & 0 & 0 & 0 & \times & \times & \times \\ 0 & 0 & 0 & 0 & 0 & \times & \times & \times \\ \times & 0 & 0 & 0 & 0 & 0 & 1 & 0 \\ 0 & \times & \times & \times & \times & 0 & 0 & 0 \\ 0 & \times & \times & \times & \times & 0 & 0 & 0 \\ 0 & \times & \times & \times & \times & 0 & 0 & 0 \end{bmatrix}. \tag{S11}$$

Here, $\times$ represents a term that can have a non-zero value based on current $\hat{\boldsymbol{m}}$ and $\hat{\boldsymbol{n}}$ orientaions and numerical values of the substrate's property tensors.

The form of $\widehat{\boldsymbol{N}}$ in Equation S11 can lead to two 4×4 null matrices along the diagonal. As eingenvalues remain invariant upon a permutation transformation of the matrix, we move row 1 to row 8 and the corresponding column 1 to column 8 in the matrix expression of S11, this results in a new matrix $\widehat{\boldsymbol{N'}}$ with a more compact form,

$$\widehat{\boldsymbol{N'}} = \begin{bmatrix} \widehat{\boldsymbol{0_4}} & \widehat{\boldsymbol{B_4}} \\ \widehat{\boldsymbol{C_4}} & \widehat{\boldsymbol{0_4}} \end{bmatrix}, \tag{S12}$$

where $\widehat{\boldsymbol{0_4}}$ is a null 4×4 matrix and $\widehat{\boldsymbol{B_4}}, \widehat{\boldsymbol{C_4}}$ are non-zero 4×4 matrices.

The eigenvalue problem of $\widehat{\boldsymbol{N'}}$ can be defined as,

$$\det\left(\widehat{\boldsymbol{N'}} - p_\alpha \widehat{\boldsymbol{I_8}}\right) = \det\left(\begin{bmatrix} -p_\alpha \widehat{\boldsymbol{I_4}} & \widehat{\boldsymbol{B_4}} \\ \widehat{\boldsymbol{C_4}} & -p_\alpha \widehat{\boldsymbol{I_4}} \end{bmatrix}\right) = 0. \tag{S13}$$

Using Schur's complement we can rewrite Equation S13 as,

$$\det\left(\widehat{\boldsymbol{N'}} - p_\alpha \widehat{\boldsymbol{I_8}}\right) = \det\left(-\widehat{\boldsymbol{I_4}}\right).\det\left(-{p_\alpha}^2 \widehat{\boldsymbol{I_4}} + \widehat{\boldsymbol{C_4}}\widehat{\boldsymbol{B_4}}\right) = 0 \tag{S14}$$

With $\det\left(\widehat{\boldsymbol{C_4}}\widehat{\boldsymbol{B_4}} - {p_\alpha}^2 \widehat{\boldsymbol{I_4}}\right) = 0$, we have an eigenvalue problem defined for the matrix product $\widehat{\boldsymbol{C_4}}\widehat{\boldsymbol{B_4}}$, which has the eigenvalue ${p_\alpha}^2$. Then,

$$\begin{aligned} &{p_\alpha}^2 = l^2 \\ &\Rightarrow p_\alpha = \pm l, \end{aligned} \tag{S15}$$

Equation S15 implies that the matrix $\widehat{\boldsymbol{N}}$ again has paired eigenvalues which differ only by a minus sign. This along with the disussion in Section 1 allows us to conclude that eigenvalues for $\widehat{\boldsymbol{N}}$ under these orientations must be either purely real or purely imaginary. Only those purely imaginary eigenvalues would be relevant to SAW states of our interest, so we will focus on $p_\alpha = \pm i|l|$ .

For a pair of eigenvalues $p_\gamma$ and $p_{\gamma+4} = -p_\gamma$ of $\widehat{\boldsymbol{N}'}$, eigenvectors $\boldsymbol{\xi}'_\gamma$ and $\boldsymbol{\xi}'_{4+\gamma}$ are $\boldsymbol{\xi}'_\gamma = (\boldsymbol{P}, \boldsymbol{Q})$ and $\boldsymbol{\xi}'_{4+\gamma} = (\boldsymbol{P}^*, \boldsymbol{Q}^*)$. $\boldsymbol{P}$, $\boldsymbol{Q}$, $\boldsymbol{P}^*$, $\boldsymbol{Q}^*$ are each a four-component vector, with * indicating the complex conjugacy between them. Since the eigenvectors of $\widehat{\boldsymbol{N}'}$ and $\widehat{\boldsymbol{N}}$ are related by the same permutation matrix that has allowed the movement of the first row and column to the last, eigenvectors of $\widehat{\boldsymbol{N}'}$ and $\widehat{\boldsymbol{N}}$ only differ by the arrangement of their elements. Given that the normalized eigenvectors of $\widehat{\boldsymbol{N}}$ are $\boldsymbol{\xi}_\alpha = \left(A_{\alpha x}, A_{\alpha y'}, A_{\alpha z'}, \phi_\alpha, L_{\alpha x}, L_{\alpha y'}, L_{\alpha z'}, D_\alpha\right)^T$, eigenvectors of the permutated $\widehat{\boldsymbol{N}'}$ now have their elements arranged as $\boldsymbol{\xi}'_\alpha = \left(A_{\alpha y'}, A_{\alpha z'}, \phi_\alpha, L_{\alpha x}, L_{\alpha y'}, L_{\alpha z'}, D_\alpha, A_{\alpha x}\right)^T$.

To understand how the eigenvectors for a given pair of imaginary eigenvalues are related, let us assume the eigenvector corresponding to $+i|l|$ be $(\boldsymbol{P}, \boldsymbol{Q})$, and $-i|l|$ be $(\boldsymbol{P}^*, \boldsymbol{Q}^*)$. Then,

$$\begin{bmatrix} \widehat{\boldsymbol{0}_4} & \widehat{\boldsymbol{B}_4} \\ \widehat{\boldsymbol{C}_4} & \widehat{\boldsymbol{0}_4} \end{bmatrix} \begin{bmatrix} \boldsymbol{P} \\ \boldsymbol{Q} \end{bmatrix} = i|l| \begin{bmatrix} \boldsymbol{P} \\ \boldsymbol{Q} \end{bmatrix}$$
$$\Rightarrow \widehat{\boldsymbol{B}_4}\boldsymbol{Q} = i|l|\boldsymbol{P}$$
$$\widehat{\boldsymbol{C}_4}\boldsymbol{P} = i|l|\boldsymbol{Q}, \tag{S16}$$

and,

$$\begin{bmatrix} \widehat{\boldsymbol{0}_4} & \widehat{\boldsymbol{B}_4} \\ \widehat{\boldsymbol{C}_4} & \widehat{\boldsymbol{0}_4} \end{bmatrix} \begin{bmatrix} \boldsymbol{P}^* \\ \boldsymbol{Q}^* \end{bmatrix} = -i|l| \begin{bmatrix} \boldsymbol{P}^* \\ \boldsymbol{Q}^* \end{bmatrix}$$
$$\Rightarrow \widehat{\boldsymbol{B}_4}\boldsymbol{Q}^* = -i|l|\boldsymbol{P}^*,$$
$$\widehat{\boldsymbol{C}_4}\boldsymbol{P}^* = -i|l|\boldsymbol{Q}^*. \tag{S17}$$

Since the eigenvectors are complex conjugates of each other,

$$\begin{bmatrix} \boldsymbol{P}^* \\ \boldsymbol{Q}^* \end{bmatrix} = \begin{bmatrix} \boldsymbol{P}e^{i2\theta_1} \\ \boldsymbol{Q}e^{i2\theta_2} \end{bmatrix}. \tag{S18}$$

Substituting Equation S18 into S17, we have,

$$\widehat{\boldsymbol{B}_4}\boldsymbol{Q} = -i|l|\boldsymbol{P}e^{-i2(\theta_2-\theta_1)},$$
$$\widehat{\boldsymbol{C}_4}\boldsymbol{P} = -i|l|\boldsymbol{Q}e^{i2(\theta_2-\theta_1)}. \tag{S19}$$

Comparing equations S19 and S16, we have,

$$e^{\pm i2(\theta_2-\theta_1)} = -1,$$
$$\Rightarrow 2(\theta_2 - \theta_1) = (2n+1)\pi \qquad n = 0,1,2 \ldots,$$
$$\Rightarrow (\theta_2 - \theta_1) = n\pi + \frac{\pi}{2} \qquad n = 0,1,2 \ldots. \tag{S20}$$

Equation S20 indicates that $A_{\alpha x}$ would have a phase difference of $(n\pi + \pi/2)$ relative to $A_{\alpha y'}, A_{\alpha z'}$, and $\phi_\alpha$. The consequence of this phase relationship will be discussed in Section 4 together with the case in Section 3.

## 3. Scenarios of non-reciprocal SAW states

Here we survey the scenarios in which neither $\widehat{\boldsymbol{m}}$ nor $\widehat{\boldsymbol{n}}$ vectors are parallel to the crystalline $x$-axis (Fig. S2). We specifically examine cases where the $x$-axis is parallel to $\widehat{\boldsymbol{m}} \times \widehat{\boldsymbol{n}}$. Under this configuration, $\widehat{\boldsymbol{m}}$ and $\widehat{\boldsymbol{n}}$ are both within the $y$-$z$ mirror plane, but could be rotated by an in-plane angle. So we can define $\widehat{\boldsymbol{m}} = y'$ and $\widehat{\boldsymbol{n}} = z'$ and use the ($x$, $y'$, $z'$) coordinate system for our $\widehat{\boldsymbol{N}}$ matrix indices, which takes the form,

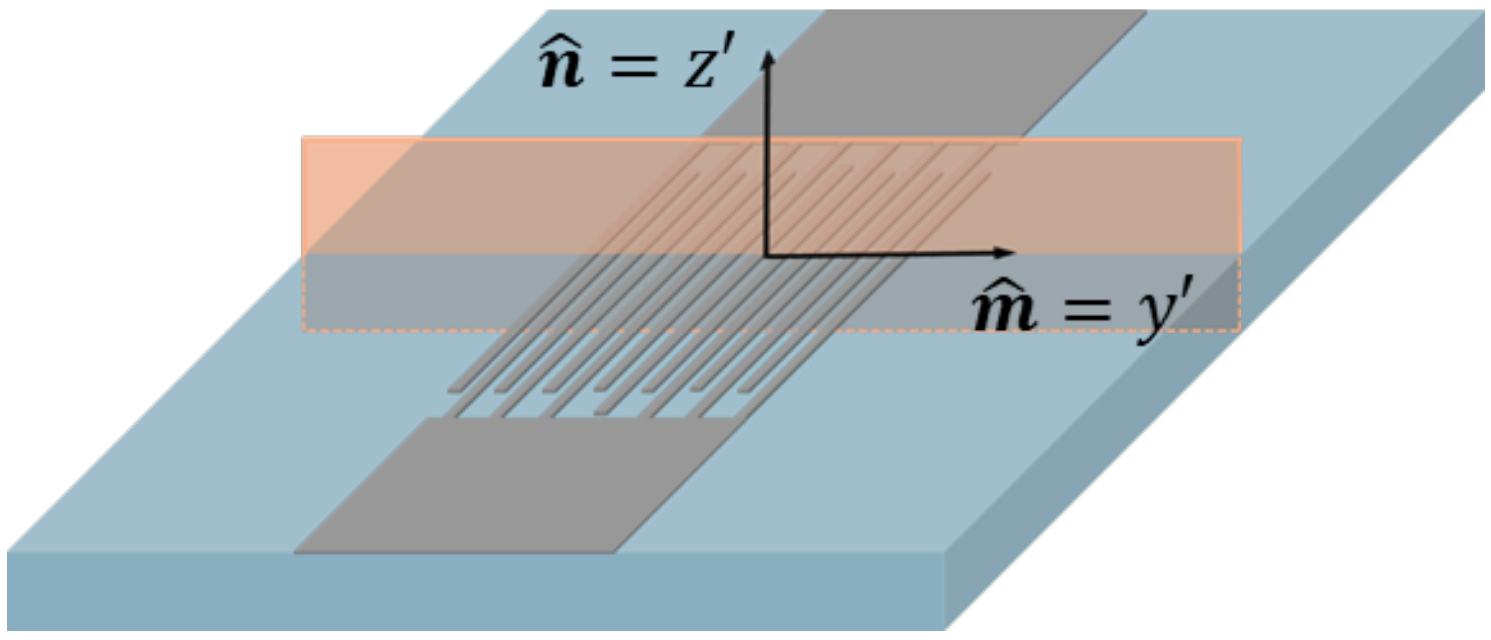

Figure S2: A schematic of non-reciprocal SAW propagation. In the specific case here, both $\hat{\boldsymbol{m}}$ and $\hat{\boldsymbol{n}}$ are confined to the mirror plane (orange) and are oriented along orthogonal axes $y'$ and $z'$.

$$\widehat{\boldsymbol{N}} = \begin{bmatrix} \times & 0 & 0 & 0 & \times & 0 & 0 & 0 \\ 0 & 0 & \times & \times & 0 & \times & \times & \times \\ 0 & 1 & \times & \times & 0 & \times & \times & \times \\ 0 & 0 & \times & \times & 0 & \times & \times & \times \\ \times & 0 & 0 & 0 & \times & 0 & 0 & 0 \\ 0 & \times & \times & \times & 0 & 0 & 1 & 0 \\ 0 & \times & \times & \times & 0 & \times & \times & \times \\ 0 & \times & \times & \times & 0 & \times & \times & \times \end{bmatrix}. \tag{S21}$$

For Equation S21, we again can make a permuation transformation to exchange the second and fifth columns (and rows). This transforms $\widehat{\boldsymbol{N}}$ into a block matrix form,

$$\widehat{\boldsymbol{N}'} = \begin{bmatrix} \times & \times & 0 & 0 & 0 & 0 & 0 & 0 \\ \times & \times & 0 & 0 & 0 & 0 & 0 & 0 \\ 0 & 0 & 0 & \times & \times & \times & \times & \times \\ 0 & 0 & 1 & \times & \times & \times & \times & \times \\ 0 & 0 & 0 & \times & \times & \times & \times & \times \\ 0 & 0 & \times & \times & \times & 0 & 1 & 0 \\ 0 & 0 & \times & \times & \times & \times & \times & \times \\ 0 & 0 & \times & \times & \times & \times & \times & \times \end{bmatrix}. \tag{S22}$$

The eigenvectors are now formatted as $\boldsymbol{\xi}'_\alpha = \left(A_{\alpha x}, L_{\alpha x}, A_{\alpha y'}, A_{\alpha z'}, \phi_\alpha, L_{\alpha y'}, L_{\alpha z'}, D_\alpha\right)^T$, which implies that $A_{\alpha x}$ and $L_{\alpha x}$ are decoupled from other components and are only dependent on each other. With both $\hat{\boldsymbol{m}}$ and $\hat{\boldsymbol{n}}$ confined to the mirror plane, this represents a horizontal shear wave along $\hat{\boldsymbol{m}} \times \hat{\boldsymbol{n}}$, which is a purly elastic wave without a dependance on the piezolelectric tensor. SAW states constructed from the rest of the components are completely confined to the mirror plane, which are the well-known piezoelectric Rayleigh waves.

The current $\widehat{\boldsymbol{N}}$ matrix form of Equation S22 cannot be further simplified as in Equations S12-S15. From Section 1, the eigenvalues thus are only complex conjugates to each other but not necessarily purely imaginary as in Equation S15. There is also no longer a phase relationship between components $(\boldsymbol{A}_{\alpha'}, \phi_\alpha, \boldsymbol{L}_\alpha, D_\alpha)$ of a given eigenvector similar to Equation S20.

For other geometries besides the scenarios in Sections 2 and 3, the SAW states will be the asymmetric type because of the projection from the current configuration of non-reciprocity.

## 4. Substrate geometry's influence on reflection coefficients

In this Section, we examine the reflection coefficients for scenarios listed in Sections 2 and 3, which provides an understanding of the reciprocity (and non-reciprocity) observed in the main text. In Section 2, we observe that for the two scenarios of reciprocity, there is a strict phase relationship between elements of the single eigenvector that define the SAW for a substrate of $3m$ point group symmetry. These eigenvectors affect the SAW reflection coefficient $r^{(\pm)}$ on IDT, which are defined as [2, 19],

$$r^{(\pm)} = R_e^{(\pm)}(\beta) + R_m^{(\pm)}(\beta)\ \frac{h_{metal}}{\lambda_0},$$

and $$R_m^{(\pm)} = \frac{2\pi\sin(\pi\beta)}{\varepsilon_\infty}\frac{\Delta v}{v}\left\{\left[\rho v^2 + 4\mu\left(1-\frac{\mu}{\lambda+2\mu}\right)\right] e_{\hat{m}}^{(\pm)^2} + [\rho v^2+\mu] e_{\hat{m}\times\hat{n}}^{(\pm)^2} + \rho v^2 e_{\hat{n}}^{(\pm)^2}\right\}, \quad \text{(S23)}$$

where $r^{(\pm)}$ is the total reflection, $R_e^{(\pm)}$ is the electrical component, $R_m^{(\pm)}$ is the mechanical component, $\beta$ the metallization ratio, $\lambda$ and $\mu$ are Lamé coefficients, $\rho$ is the density of IDT material, $h_{metal}$ is the thickness of the IDT, $\lambda_0$ is the wavelength of the SAW, and $e_i^{(\pm)}$ are the components of the vector $\boldsymbol{e}^{(\pm)} = \boldsymbol{A}^{(\pm)}/\phi^{(\pm)}$, where $i = \hat{m}, \hat{m}\times\hat{n}$ or $\hat{n}$.

The consequence of Equation S20 to Equation S23 is that components of $\boldsymbol{e}^{(\pm)}$ along two opposite directions must have the same magnitude, but also a phase difference of either 0 or $(n\pi + \pi/2)$. Then, the quantity $(\boldsymbol{e}^{(\pm)})^2$ would be purely real. According to Ref. [2], this implies that there cannot be any NSPUDT effect under these geometries. However, in the non-reciprocal case, the phase relation between $\boldsymbol{A}^{(\pm)}$ and $\phi^{(\pm)}$ is arbitrary, so it allows $(\boldsymbol{e}^{(\pm)})^2$ to be complex. Thus, these geometries allow for the amplification of non-reciprocity (asymmetry) between SAW states, based on multiple internal reflections of SAW within the same IDT. Our experimental results in the main text confirm that the asymmetry grows with both the height and the number of fingers of IDT, as they both promote internal reflections within the IDT.